\documentstyle[twocolumn,epsf]{jpsj}

\def\runtitle{Enhancement of Pairing Correlation and Spin Gap
through Suppression of Single-Particle Dispersion in One-Dimensional Models
}
\def\runauthor{Masanori {\sc Kohno} and Masatoshi {\sc Imada}}

\title
{
Enhancement of Pairing Correlation and Spin Gap 
through Suppression of Single-Particle Dispersion in One-Dimensional Models
}

\author
{ 
Masanori {\sc Kohno} and Masatoshi {\sc Imada}$^a$
}

\inst
{
Mitsubishi Research Institute, Inc., Otemachi, Chiyoda-ku, Tokyo 100-8141, \\
and \\
$^{a}$Institute for Solid State Physics, University of Tokyo, Roppongi, Minato-ku, 
Tokyo 106-8666
}

\recdate
{
\today
}

\abst
{
We investigate the effects of suppression of single-particle dispersion near the Fermi level 
on the spin gap and the singlet-pairing correlation by using the exact diagonalization method 
for finite-size systems. 
We consider strongly correlated one-dimensional models, which have flat band dispersions 
near wave number $k=\pi/2$, if the interactions are switched off. 
Our results for strongly correlated models show that the spin gap region expands 
as the single-particle dispersion becomes flatter. The region where
the singlet-pairing correlation is the most 
dominant also expands in models with flatter band dispersions. 
Based on our numerical results, we propose a pairing mechanism induced by the 
flat-band dispersion.
}

\kword
{
flat band, single-particle dispersion, spin gap, superconductivity
}

\begin{document}
\sloppy
\maketitle

One of the characteristic features of hole-doped high-$T_{\rm c}$ cuprates is the emergence 
of flat single-particle dispersion near the X points [i.e. $k=(\pi,0),(0,\pi)$].\cite{Abrikosov,Gofron,Dessau} 
This unusual flatness of the dispersion relation is ascribed to strong electronic correlations.\cite{Dagotto,Bulut} 
A recent numerical study on the two-dimensional (2D) Hubbard 
model showed that the flat single-particle dispersion near the X points is related to the universality class of the Mott transition.\cite{AssaadEur} 
Pairing instabilities of systems under this universality class were also studied,\cite{AssaadIS96,AssaadIS97,Assaadk4,Tsunetsugu} 
where the dynamical exponent $z$ changes from four to two for the 2D Hubbard and $t$-$J$ 
models when a two-particle hopping term is switched on.
This shows potential instability of flat single-particle dispersions towards 
singlet-pair formation, if two-particle hopping processes are introduced.
\par
Taking this into account, we consider a possible scenario for enhancing the pairing instability: 
In strongly correlated systems, the two-particle hopping process can mainly be controlled by the 
incoherent part of charge excitations independently of the coherent part of single-particle excitations, 
while such a coherent single-particle process may rather suppress the pairing susceptibility through damping 
and pair breaking processes. Therefore, if only the single-particle dispersion near the Fermi level is suppressed,
the two-particle hopping process determined from the dispersive band far from the Fermi level would enhance the pairing instabilities.\cite{ImadaKohno} 
\par
In this letter, in order to confirm the occurrence of this scenario in a simplified situation, we introduce 
one-dimensional (1D) models that have flat single-particle dispersion near half filling.\cite{Comment1} 
We clarify how the singlet-pairing correlation and the spin gap behave in the parameter 
space of our model.
\par
\begin{figure}
\begin{center}
\epsfxsize=3.2in \leavevmode \epsfbox{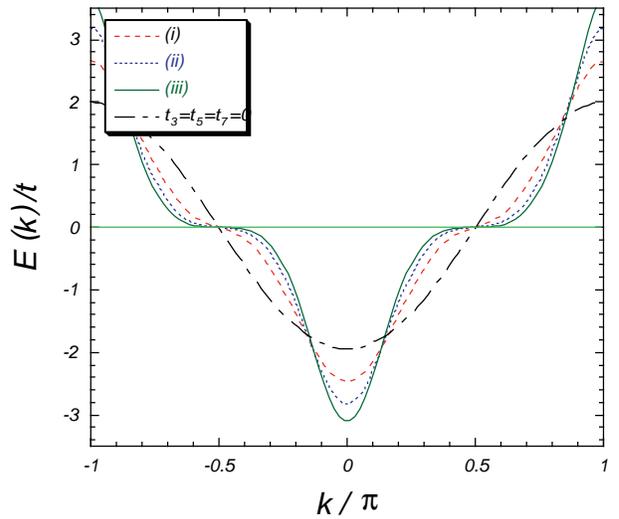}
\end{center}
\caption{Dispersion relations of noninteracting models with long-range hopping terms, 
where $t_3/t, t_5/t, t_7/t$ take values of (i), (ii), (iii) in the text. The horizontal line corresponds 
to the Fermi level at half-filling. For comparison, the dispersion relation of the 
noninteracting model without long-range hopping terms is also plotted.}
\label{disp_non}
\end{figure}

We consider the Hubbard model with long-range hopping terms defined 
by the following Hamiltonian:
\begin{subequations}
\begin{eqnarray}
   {\cal H}_{0}&\equiv&{\cal H}_{\rm Hub}+{\cal H}_{\rm lrh},\\
   {\cal H}_{\rm Hub}&\!\equiv\!&-t\!\sum_{i,\sigma}(c^{\dagger}_{i\sigma}
c_{i\!+\!1\sigma}\!+\!c^{\dagger}_{i\!+\!1\sigma}c_{i\sigma})
\!+\!U\!\sum_{i}\!n_{i\uparrow}n_{i\downarrow},\\
   {\cal H}_{\rm lrh}&\equiv&-t_3\sum_{i,\sigma}(c^{\dagger}_{i\sigma}c_{i+3\sigma}+
c^{\dagger}_{i+3\sigma}c_{i\sigma})\nonumber\\
&&
-t_5\sum_{i,\sigma}(c^{\dagger}_{i\sigma}c_{i+5\sigma}+c^{\dagger}_{i+5\sigma}c_{i\sigma})\nonumber\\
&&
-t_7\sum_{i,\sigma}(c^{\dagger}_{i\sigma}c_{i+7\sigma}+c^{\dagger}_{i+7\sigma}c_{i\sigma}).
\end{eqnarray}
\end{subequations}
We take hopping amplitude $t_i$ ($i=3,5,7$) of the long-range hopping terms as (i) $t_3/t=1/3, 
t_5/t=t_7/t=0$, (ii) $t_3/t=0.5, t_5/t=0.1, t_7/t=0$, and (iii) $t_3/t=0.6, t_5/t=0.2, t_7/t=1/35$ 
in order to realize flat dispersion near half filling in the noninteracting case. With this tuning, 
the noninteracting dispersion around $q_0=\pm\pi/2$ is suppressed up to $(q-q_0)^{\nu}$, 
where $\nu=2,4$ and $6$ for (i), (ii) and (iii), respectively. The dispersion relations of 
these noninteracting models are shown in Fig. \ref{disp_non}. The region of flat-band dispersion 
near half filling is allowed to expand further as the hopping range becomes longer. 
Taking the limit of strong correlation (i.e. $U\rightarrow\infty$), we obtain the following 
effective Hamiltonian:
\begin{subequations}
\begin{eqnarray}
   {\cal H}&\equiv&{\cal H}_{t-J}+{\cal H}_{\rm 3-site}+{\tilde {\cal H}}_{\rm lrh}+{\cal H}_{{\rm long}J},\\
   {\cal H}_{t-J}&\equiv&-t\sum_{i,\sigma}({\tilde c}^{\dagger}_{i\sigma}{\tilde c}_{i+1\sigma}
+{\tilde c}^{\dagger}_{i+1\sigma}{\tilde c}_{i\sigma})\nonumber\\
&&+J\sum_{i}({\mbox{\boldmath S}}_i\cdot{\mbox{\boldmath S}}_{i+1}-\frac{1}{4}n_{i}n_{i+1}),\\
   {\cal H}_{\rm 3-site}&\equiv&-\frac{J}{4}\sum_{i,\sigma}({\tilde c}^{\dagger}_{i-1,\sigma}
n_{i,-\sigma}{\tilde c}_{i+1,\sigma}\nonumber\\
&&-{\tilde c}^{\dagger}_{i-1,\sigma}{\tilde c}^{\dagger}_{i,-\sigma}
{\tilde c}_{i,\sigma}{\tilde c}_{i+1,-\sigma}+{\rm H.c.}),\\
   {\tilde {\cal H}}_{\rm lrh}&\equiv&
-t_3\sum_{i,\sigma}({\tilde c}^{\dagger}_{i\sigma}{\tilde c}_{i+3\sigma}+
{\tilde c}^{\dagger}_{i+3\sigma}{\tilde c}_{i\sigma})\nonumber\\
&&
-t_5\sum_{i,\sigma}({\tilde c}^{\dagger}_{i\sigma}{\tilde c}_{i+5\sigma}
+{\tilde c}^{\dagger}_{i+5\sigma}{\tilde c}_{i\sigma})\nonumber\\
&&
-t_7\sum_{i,\sigma}({\tilde c}^{\dagger}_{i\sigma}{\tilde c}_{i+7\sigma}
+{\tilde c}^{\dagger}_{i+7\sigma}{\tilde c}_{i\sigma}).
\end{eqnarray}
\end{subequations}
Here, ${\cal H}_{{\rm long}J}$ denotes the spin exchange and three-site terms originating 
from long-range hopping.
In this letter, we mainly study the model defined by the Hamiltonian ${\cal H}_{tJ3{\rm l}}\equiv
{\cal H}_{t-J}+{\cal H}_{\rm 3-site}+{\tilde {\cal H}}_{\rm lrh}$ for simplicity. 
Later, we discuss the effects of ${\cal H}_{{\rm long}J}$ briefly. 
The operator ${\tilde c}^{\dagger}_{i\sigma}$ creates an electron with spin $\sigma$ 
at site $i$ in the subspace without double occupancy. 
The operators ${\mbox{\boldmath S}}_i$ and $n_i$ are the spin and number operators 
at site $i$, respectively. Hereafter, the chain length and the number of electrons are 
denoted by $L$ and $N_{\rm e}$, respectively. 
We set $t=1$ as the energy unit.
\par
We call models defined by ${\cal H}_{tJ3{\rm l}}$ with (i) $t_3=1/3, 
t_5=t_7=0$, (ii) $t_3=0.5, t_5=0.1, t_7=0$, (iii) $t_3=0.6, t_5=0.2, 
t_7=1/35$, and (iv) $t_3=t_5=t_7=0$ as tJ3t3, tJ3t5, tJ3t7 and tJ3, 
respectively. The tJ3 model is 
nothing but the effective model of the strong-coupling limit of the Hubbard model up to the order 
$t^2/U$.\cite{tJ3drv} The ground-state properties of this model were investigated in detail by Ammon, Troyer and Tsunetsugu.\cite{Ammon}
\par
The aim of this study is to demonstrate that the suppression of single-particle hopping processes 
enhances singlet pair correlations. We show this enhancement by studying the effects of the above 
long-range hopping terms ${\tilde {\cal H}}_{\rm lrh}$ on the spin gap and the singlet-pairing 
correlation.
\par
\begin{figure}
\begin{center}
\epsfxsize=3.2in \leavevmode \epsfbox{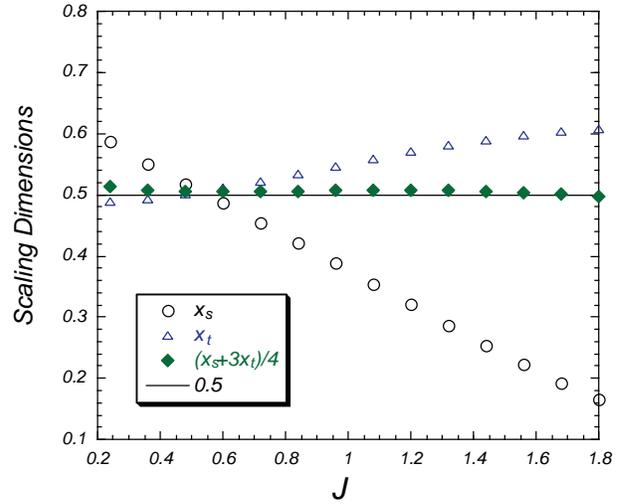}
\end{center}
\caption{Scaling dimension of the lowest singlet excitation ($x_{\rm s}$) 
and that of triplet excitation ($x_{\rm t}$) at $k=2k_{\rm F}$ for the
tJ3t7 model at quarter-filling in a 16-site cluster. 
We also plot the quantity $x_{\rm r}\equiv (x_{\rm s}+3x_{\rm t})/4$.}
\label{scaling}
\end{figure}
\begin{figure}
\begin{center}
\epsfxsize=3.2in \leavevmode \epsfbox{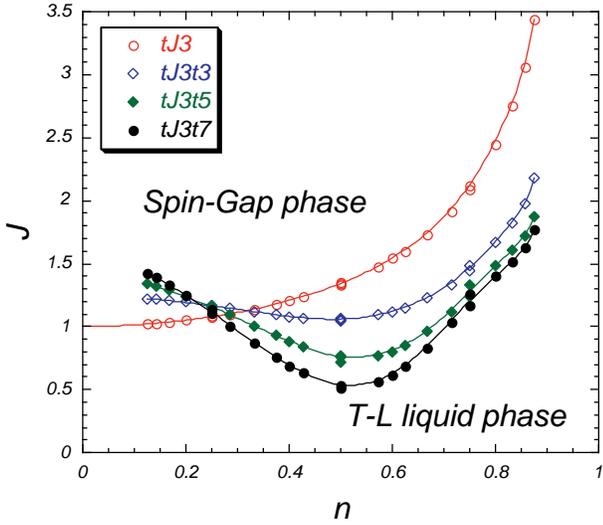}
\end{center}
\caption{Phase boundary between the Tomonaga-Luttinger (TL) liquid phase and 
the spin-gap (SG) phase for the tJ3, tJ3t3, tJ3t5 and tJ3t7 models. The symbols are results for clusters 
of length $L=8,10,12,14$ and $16$.}
\label{phSG}
\end{figure}

First, we show numerical results on the phase boundary between the Tomonaga-Luttinger (TL) liquid 
phase\cite{Tomonaga,Haldane} and the spin-gap (SG) phase. We employ the singlet-triplet 
level crossing method
\cite{JulienHaldane,Affleck,Nomura,Nakamura_tJ,Nakamura_tJJ,Nakamura_eHub,Nakamura_rev} 
in order to determine the phase boundary. The effectiveness of this method has been demonstrated 
through the applications to 1D spin systems\cite{JulienHaldane,Affleck,Nomura} 
and electron systems.\cite{Nakamura_tJ,Nakamura_tJJ,Nakamura_eHub,Nakamura_rev} 
Using this method, we determine the critical point with high precision,\cite{JulienHaldane,Nomura} 
since the finite-size correction is only due to irrelevant fields with scaling dimension $x=4$.\cite{Cardy} This method is based on the assumption that the universality class of the phase transition is described by the level $k=1$ SU(2) Wess-Zumino-Witten (WZW) model.\cite{Affleck} 
In order to check the applicability of this method to our models, we have calculated the scaling 
dimension of the lowest singlet excitation ($x_{\rm s}$) and that of triplet excitation ($x_{\rm t}$) 
at wave number $k=2k_{\rm F}$. According to the renormalization-group analysis, the leading 
finite-size corrections of $x_{\rm s}$ and $x_{\rm t}$ near the critical point are written as\cite{Nomura,Nakamura_tJ}
\begin{subequations}
\begin{eqnarray}
x_{\rm s}&=&\frac{1}{2}+\frac{3}{4}\frac{y_0}{y_0\log L+1},\\
x_{\rm t}&=&\frac{1}{2}-\frac{1}{4}\frac{y_0}{y_0\log L+1},
\end{eqnarray}
\end{subequations}
where $y_0$ is a coupling constant. As shown in Fig. \ref{scaling}, the value of 
$x_{\rm r}\equiv (x_{\rm s}+3x_{\rm t})/4$ is very close to 1/2 and independent of the value of 
parameter $J$. This indicates that the low-energy physics of our models is described as that of the 
Tomonaga-Luttinger liquid, and that the transition belongs to the universality class of the $k=1$ 
SU(2) WZW model. In fact, as shown in Fig. \ref{phSG}, the phase boundaries in clusters of length $L=8,10,12,14$ and $16$ are scaled in single lines. This suggests that the size dependence 
is very small.
\par
Figure \ref{phSG} shows the phase boundaries between the TL liquid phase and the SG phase for the 
tJ3, tJ3t3, tJ3t5 and tJ3t7 models. Clearly, the region of the SG phase expands further 
near half filling as the hopping range becomes longer. In other words, the suppression of the Fermi velocity 
enlarges the region of the spin-gap phase. This may be explained as follows: 
Generally, low-energy properties of fermionic systems are mainly determined by scattering 
processes near the Fermi level. The suppression of the Fermi velocity would have similar effects 
as the reduction of hopping amplitude $t$ on the low-energy physics, because the Fermi velocity 
is roughly proportional to $t$.  The long-range hopping terms, which suppress the Fermi velocity, 
effectively decrease $t$, while they do not alter $J$ and the three-site terms. Then, they increase 
the effective $J/t$ for the tJ3 model. In the large $J/t$ regime for the tJ3 model, the spin excitation 
has an energy gap, because of the large two-particle hopping processes such as spin exchange 
and three-site hopping.\cite{Ammon} Hence, the suppression of single-particle hopping processes 
due to the long-range hopping terms results in the enlargement of the spin-gap phase.
\par
It should be noted that the spin-gap region extends only in the finite-doping regime, not at half filling. This is analogous to the behavior of 
high-$T_{\rm c}$ cuprates and different from that of the dimerized 
model,\cite{dimer} the $t$-$J$-$J^{\prime}$ 
model\cite{Ogata_tJJ,Sano_tJJ,Nakamura_tJJ} and the $t$-$J$ ladder model,\cite{Troyer,Hayward,Sano_lad} 
where the spin gap opens not only in the finite-doping regime, but also at half filling. This difference results in different classes of proximity effects 
from Mott insulators and hence different pairing mechanisms. 
Our study provides some insight on this issue.
\par
\begin{figure}
\begin{center}
\epsfxsize=3.2in \leavevmode \epsfbox{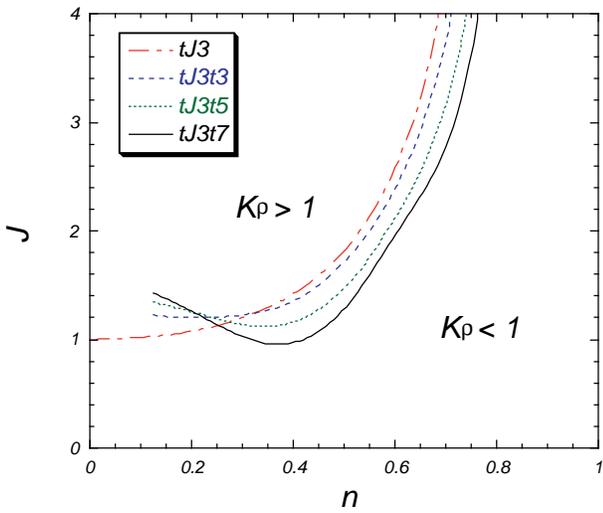}
\end{center}
\caption{Contour lines of $K_{\rho}=1$ for the tJ3, tJ3t3, tJ3t5 and tJ3t7 models in a 16-site cluster. }
\label{Kr1}
\end{figure}

Next, we consider correlation functions. Generally, in 1D interacting systems, the long-range behavior 
of correlation functions is described by a single correlation exponent 
$K_{\rho}$.\cite{Haldane,Frahm} 
The correlation exponent $K_{\rho}$ is obtained as 
\begin{equation}
K_{\rho}=\pi\sqrt{\frac{\sigma_0n^2\kappa}{2}},\nonumber
\end{equation}
where Drude weight of the ac conductivity $\sigma_0$ and compressibility $\kappa$ 
in finite-size systems are respectively defined\cite{Kohn,Shastry,Ogata_tJ} as
\begin{subequations}
\begin{eqnarray}
\sigma_0&\equiv&\frac{L}{2}\frac{\partial^2 E_0(\phi)}{\partial \phi^2}|_{\phi=0},\\
\kappa&\equiv&\!\frac{L}{N_{\rm e}^2}\!\frac{4}{E_0(L\!,\!N_{\rm e}\!+\!2)\!+\!E_0(L\!,\!N_{\rm e}\!-\!2)
\!-\!2E_0(L\!,\!N_{\rm e})}\!.
\end{eqnarray}
\end{subequations}
Here, $E_0(\phi)$ denotes the ground-state energy of the system with twisted 
boundary conditions with phase factor $\phi$. $E_0(L,N_{\rm e})$ denotes the 
ground-state energy of the $N_{\rm e}$-electron system of length $L$, and $n$ is the electron 
density defined by $n\equiv N_{\rm e}/L$. The long-range behavior of the singlet-pairing correlation 
function $P(r)$ is expressed in terms of $K_{\rho}$ as $P(r)\propto1/r^{1+1/K_{\rho}}$ 
if the spin excitation is gapless, while $P(r)\propto1/r^{1/K_{\rho}}$ in the spin gap phase. 
If $K_{\rho}$ is larger than one, the singlet-pairing correlation is the most dominant among other 
correlation functions. We show the contour lines of $K_{\rho}=1$ in Fig. \ref{Kr1}. 
The region of $K_{\rho}>1$ expands as the hopping range becomes longer. 
This means that the suppression of single-particle hopping enhances the singlet-pairing correlation. 
\par
\begin{figure}
\begin{center}
\epsfxsize=3.2in \leavevmode \epsfbox{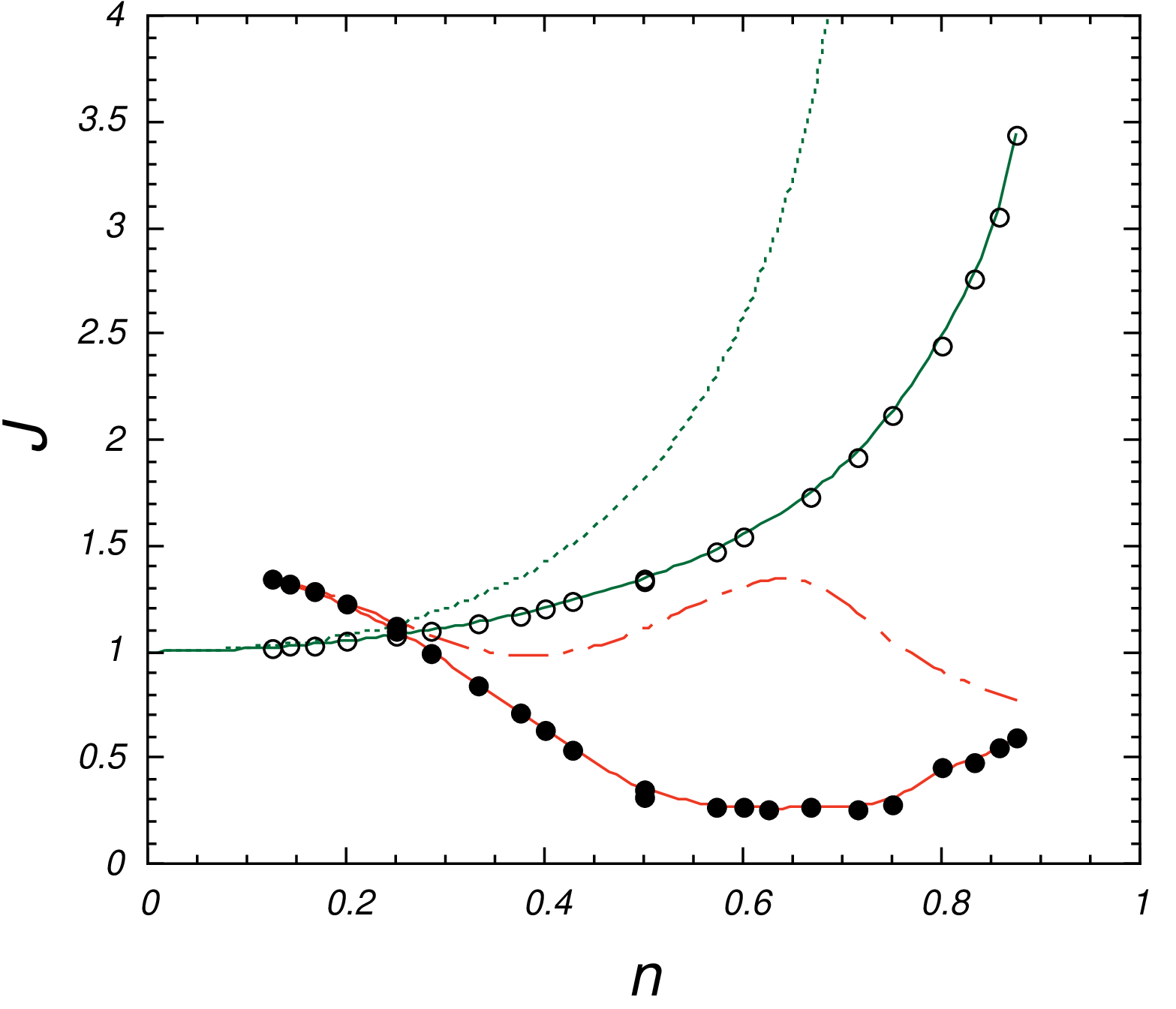}
\end{center}
\caption{Spin-gap phase boundary and the contour line of $K_{\rho}=1$ for the tJ3t7 model 
with the long-range spin exchange term. The spin-gap phase boundary and the contour line of $K_{\rho}=1$ for this model are represented by solid circles 
and a dash-dotted line, respectively. 
For comparison, we also plot those of the tJ3 model. The spin-gap phase boundary and the contour line of 
$K_{\rho}=1$ for the tJ3 model are represented by open circles and a dotted line, respectively.}
\label{longJ}
\end{figure}

Here, we briefly refer to the effects of the neglected terms in ${\cal H}_{{\rm long}J}$.\cite{ImadaKohno} 
We consider the system defined by the following Hamiltonian:
\begin{subequations}
\begin{eqnarray}
   {\cal H}&\equiv&{\cal H}_{t-J}+{\cal H}_{\rm 3-site}+{\tilde {\cal H}}_{\rm lrh}
+{\cal H}_{{\rm lex}J},\\
   {\cal H}_{{\rm lex}J}&\equiv&
J_3\sum_{i}({\mbox{\boldmath S}}_i\cdot{\mbox{\boldmath S}}_{i+3}-\frac{1}{4}n_{i}n_{i+3})\nonumber\\
&&+J_5\sum_{i}({\mbox{\boldmath S}}_i\cdot{\mbox{\boldmath S}}_{i+5}
-\frac{1}{4}n_{i}n_{i+5})\nonumber\\
&&+J_7\sum_{i}({\mbox{\boldmath S}}_i\cdot{\mbox{\boldmath S}}_{i+7}-\frac{1}{4}n_{i}n_{i+7}),
\end{eqnarray}
\end{subequations}
where coupling constant $J_j$ is defined as $J_j\equiv J\times t_j^2$ ($j=3,5$ and $7$).
The spin-gap phase boundary and the contour line of $K_{\rho}=1$ for the above model 
with $t_3=0.6, t_5=0.2, t_7=1/35$ are shown in Fig. \ref{longJ}. This shows that the long-range 
spin exchange interaction also enlarges the spin-gap region and the region 
of $K_{\rho}>1$. The region of singlet pairs may be determined mainly by the following two 
factors: One is two-particle hopping processes, which enhance the coherent motion of electron pairs 
and prevent phase separation, and the other is antiferromagnetic fluctuations, which induce 
attractive interactions between up and down spins. For the $t$-$J$ model in the large $J$ regime, 
the spin-gap phase exists, but the antiferromagnetic fluctuations are so large that the spin-gap 
phase is confined in a narrow parameter regime due to instability resulting from phase separation.\cite{Ogata_tJ,Nakamura_tJ} 
For the tJ3 model, single- and two-particle hopping 
processes are strong enough to prevent phase separation.\cite{Ammon} However, since the larger 
single-particle hopping is accompanied by the three-site term, it requires a large $J$ to reach the 
region of $K_{\rho}>1$. 
The numerical results shown in Fig. \ref{longJ} indicate that the suppressed single-particle coherence 
together with preserved two-particle hopping and finite-range spin exchanges contribute to enhancing both 
the effective attractive interactions between up and down spins and the coherent pair motion.
\par
Here, we consider possible realizations that may be described by the above models. An example is 
a three-chain system: The central chain is composed of atoms with $d$-orbitals where the Fermi level 
is located. The other two chains are composed of atoms with $\pi$-orbitals. These $\pi$-orbitals are designed 
to lie near the Fermi level and to overlap with the $d$-orbitals
of the central chain. In such a system, electrons on $d$-orbitals can hop to a distant site 
by virtual hopping processes through the side chains. The low-energy physics may be described by 
the tJ3t3 model if the parameter is tuned. 
Another is a spiral chain: The orbitals of third neighbor atoms in a chain may 
overlap, if the chain has a spiral structure. 
The overlap between orbitals may produce single-particle hopping processes 
between distant sites. 
In such a system, the low-energy physics may be described by the tJ3t3 model. 
\par
In summary, we have demonstrated that the suppression of the single-particle hopping processes 
near the Fermi level enlarges both the spin-gap region and the region for the dominance of the pairing. 
Based on the present results, we propose a possible scenario for a pairing mechanism: 
If the suppression of the single-particle hopping is sufficient, pairing correlations would be 
enhanced, which may lead to superconductivity.
\par
Although the models investigated by us are specific and somewhat artificial, the underlying physics 
would be universal. The concept of flat-band induced pairing instability may be appropriate not only 
for one-dimensional models with long-range hopping terms, but also for models with short-range hopping 
terms in higher dimensions, if the single-particle hopping processes are sufficiently suppressed near the Fermi level close to the Mott insulator.

The authors would like to thank M. Ishikawa for discussions. 
One of the authors (M.K.) thanks M. Nakamura for helpful comments on the singlet-triplet level crossing 
method. The exact diagonalization program is partly based on the subroutine package "TITPACK Ver.2" 
coded by H. Nishimori. This work is financially supported by a Grant-in-Aid for "Research for the 
Future" Program from the Japan Society for the Promotion of Science under the project 
JSPS-RFTF97P01103.

\newpage

\begin{thebibliography}{99}
\bibitem{Comment1} In strongly correlated systems near half filling, 
a two-particle process can be controlled rather independently of a single-particle 
process. Thus, the effects of flat band dispersion would become pronounced 
near half filling.
\label{Comment1}
\end{thebibliography}
\end{document}